# Structural properties and cation ordering in layered hexagonal $Ca_xCoO_2$

H.X. Yang*, Y.G. Shi, X. Liu, R.J. Xiao, H.F. Tian and J.Q. Li

Beijing National Laboratory for Condensed Matter Physics, Institute of Physics, Chinese Academy of Sciences, Beijing 100080, China

A series of $Ca_xCoO_2$ ($0.15 \leq x \leq 0.40$) materials have been prepared by means of ion exchange reaction from $Na_xCoO_2$. Transmission electron microscopy (TEM) measurements revealed a rich variety of structural phenomena resulting from cation ordering, structural distortion and twinning. Systematical structural analysis, in combination with the experimental data of $Na_xCoO_2$ ($0.15 \leq x \leq 0.8$) and $Sr_xCoO_2$ ($1.5 \leq x \leq 0.4$) systems, suggests that there are two common well-defined cation ordered states corresponding respectively to the orthorhombic superstructure at around $x = 1/2$ and the $3^{1/2}a \times 3^{1/2}a$ superstructure at around $x = 1/3$ in this kind of system. Multiple ordered states, phase separation, and incommensurate structural modulations commonly appear in the materials with $0.33 < x < 0.5$. The TEM observations also reveals an additional periodic structural distortion with $\mathbf{q}_2 = \mathbf{a}^* / 2$ in materials for $x \leq 0.35$. This structural modulation also appears in the remarkable superconducting phase $Na_{0.33}CoO_2$ $1.3H_2O$.

PACS numbers: 61.66. Fn, 61.14.Lj, 71.27.+a, 61.44.Fw

Keywords: $Ca_xCoO_2$; Cobalt oxides, Electron diffraction, Superstructure, Cation ordering

Layered deintercalatable alkali metal oxides, such as $Li_xCoO_2$ and $Na_xCoO_2$ ($0 < x < 1.0$), have been the major subject of intense research activities in the past years owing to their potential technological applications and remarkable physical properties [1-3]. The noteworthy features of these materials, from both structural and chemical point of views, are that the cation content and crystal structure can vary over a wide range by ionic deintercalation and exchange, and, moreover, the physical properties are profoundly affected by the cation concentration and vacancy ordering [4-5]. Recently, experimental and theoretical investigations have paid special attention to intercalated cation/vacancy ordering in the layered cobalt oxides [6-14]. First principle calculations demonstrated that a variety of possible ordered states can be stable at different intercalated cation contents, such as $3^{1/2}a \times 3^{1/2}a$ superstructure at x = 1/3 (or 2/3) and other ordered states at x = 1/2, 1/4, 1/5, et al [6, 7]. Previously, TEM observations on superstructures in correlation with the cation ordering have demonstrated several ordered states in $Li_xCoO_2$ and $Na_xCoO_2$ materials [8, 9, 12]. However, it is also noted that the electron beam radiation during TEM observation could severely change cation arrangements among the $CO_2$ sheets due to the high mobility of either Na or Li ions in this layered structure [8-14].

Precise information of local structure is a key issue to understand the correlation between structure and physical properties. For instance, Na atoms in $Na_{0.5}CoO_2$ crystallize in a well-defined zigzag ordered pattern yielding an orthorhombic structure in which low-temperature charge ordering is observed [11-13]. $M_xCoO_2$ (M = Sr, Ca) compounds are expected to be more stable than their analogous $Na_xCoO_2$ materials due to the greater electrostatic interaction between divalent ions and the negative $CoO_2$ layer, which makes $(Sr, Ca)_xCoO_2$ more suitable for TEM observations [15,17]. In present study, considering the ionic radius of $Ca^{2+}$ (1.0 Å) is comparable to $Na^+$ (1.02 Å), we have performed an extensive study on $Ca_xCoO_2$ (0.15 ≤ x ≤ 0.4) materials and expect to obtain systematic data for the understanding of local structure properties in this layered system.

Polycrystalline materials with nominal compositions of $Ca_xCoO_2$ (0.15 ≤ x ≤ 0.4) were prepared by the low-temperature ion exchange reaction from the γ-$Na_xCoO_2$ (0.33 ≤ x ≤ 0.8) precursor prepared by conventional solid-state reaction or by sodium deintercalation of $Na_{0.75}CoO_2$ [5]. The ion exchange process was carried out by using the modified Cushing-Wiley method [15]. An amount of $Na_xCoO_2$ was mixed thoroughly with 10% molar excess anhydrous $Ca(NO_3)_2$ powder, then heated at 310 °C for two days in air, the mixture was grinded repeatedly during this process. The final products are washed by distilled water. The metal ratios in the products were determined by inductively coupled plasma (ICP) analysis. Powder XRD was performed by employing a RIGAKU x-ray diffractometer. Most specimens for TEM observations were prepared simply by crushing the bulk material into fine fragments, which were then supported on a copper grid coated with a thin carbon film. We also prepared some samples by pressing polycrystalline into pellets, which then were polished mechanically with a Gatan polisher to a thickness of around 50μm and then ion-milled by a Gatan-691 PIPS ion miller. The TEM investigations were performed on a TECNAI F20 operating at a voltage of 200 kV.

Table 1 lists the experimental lattice parameters for the materials with nominal compositions of $Ca_xCoO_2$ (0.15 ≤ x ≤ 0.4) and the relevant precursors $Na_xCoO_2$ (x = 0.5, 0.7). It is demonstrated that the lattice parameter **a** and **c** change only gradually with Ca content and this is quite different from $Na_xCoO_2$ as illustrated in table 1 for x = 0.5 and 0.7. $Na_xCoO_2$ materials in general show evident structural evolution, in particular for the **c** parameter, with sodium content. This kind of change is directly connected with the coupling between the $CoO_2$ sheet and the intercalated (Na or Ca) layers. Fig. 1 shows the XRD patterns obtained from materials with nominal compositions of $Ca_{0.25}CoO_2$ and $Na_{0.5}CoO_2$. The structural transformation from orthorhombic $Na_{0.5}CoO_2$ to hexagonal $Ca_{0.25}CoO_2$ can be clearly recognized. This fact directly confirms that the orthorhombic structural feature in $Na_{0.5}CoO_2$ arising from the zigzag Na ordering is totally different in the $Ca_{0.25}CoO_2$ as discussed in the following context.

In order to get a better understanding of the common structural properties in $M_xCoO_2$ (M = Na, Sr, Ca) systems, we have systematically analyzed superstructures observed in $Na_xCoO_2$ (0.33 ≤ x ≤ 0.8) [12], $Sr_xCoO_2$ (0.25 ≤ x ≤ 0.4) [17] and $Ca_xCoO_2$ (0.15 ≤ x ≤ 0.4) materials. It should be mentioned that, due to the high mobility of Na cations, numerous observed superstructures in $Na_xCoO_2$ arise artificially from the electron beam irradiation during the TEM observation. On the other hand, the $Ca_xCoO_2$ and $Sr_xCoO_2$ materials are found to be rather stable and no beam induced structural change is observed during our TEM observations. Based on our experimental results for $Ca_xCoO_2$ and the previous experimental and theoretical analysis on the $Sr_xCoO_2$ and $Na_xCoO_2$ materials [6, 7, 12, 17], we show a simplified phase diagram in fig. 2 (a) illustrating the presence of ordered structures along with cation content. It is demonstrated that two kinds of well-defined superstructure phases exist respectively at around x = 1/2 and 1/3 (fig. 2 (b) and fig 2 (c)), i.e. an orthorhombic type superstructure as systematically discussed in $Na_{0.5}CoO_2$ and a $3^{1/2}a \times 3^{1/2}a$ superstructure observed in our previous study of $Sr_{0.35}CoO_2$. In general, the notable structural features revealed in these systems are the presence of the complex structural modulations and remarkable phase separation, especially for samples with x ranging from 1/3 to 1/2 [12, 17]. In the $Na_{0.7}CoO_2$ materials, the Na cations were found to occupy two possible positions $P_1$ and $P_2$ governed by the space between the two adjacent oxygen planes. Both sites are trigonal prismatic sites, but one trigonal prism shares edges with adjacent $CoO_6$ octahedral, the other trigonal prism shares faces with adjacent $CoO_6$ octahedral [13]. In the orthorhombic structure, a cation zig-zig chain structure with 50% occupation on either $P_1$ or $P_2$ site has been proposed and further confirmed by the neutron diffraction and TEM studies on $Na_{0.5}CoO_2$ [6, 11, 12, 13]. The incommensurability observed in this layered system is proposed to originate from an insertion of an extra vacancy plane within the zig-zag type of intercalated atomic ordering, which can yield visible space and orientation anomalies in the electron diffraction patterns [12, 13]. The $3^{1/2}a \times 3^{1/2}a$ superstructure can be explained by the cation ordering at $P_2$ sites, which has been well verified in the our HRTEM study of $Sr_{0.35}CoO_2$ in agreement with the data of first principle calculations based on a maximum distance between the cation atoms [6, 17]. Moreover, certain interesting structural features, such as multi-modulated structures, appear commonly in the range of x ≤ 0.35 in correlation with cation ordering and related structural distortions. For instance, a structural distortion caused by atomic shifts primarily along the c axis direction has been observed in $Sr_{0.35}CoO_2$, this modulation has a wave vector of $q_2 = a^*/2$ and becomes gradually stronger with lowering cation content [17].

In addition to the well-defined superstructures, our TEM experimental investigations on the $Ca_xCoO_2$ samples with x in the range of 0.3 to 0.4 in general reveal complicated Ca-ordered states and phase segregation as similarly discussed in the $Na_xCoO_2$ system in our previous works [10, 13]. Figure 2 (d) displays a dark-field TEM image of a $Ca_{0.4}CoO_2$ crystal, indicating the presence of phase separation in the examined area. This image was obtained by using the superstructure spots, thus the evident contrast alternation directly indicates the change of Ca ordering from one area to another. Actually, this kind of domain structure is often seen in incommensurate or nearly commensurate modulated systems. Electron diffraction observations occasionally reveal the well-defined orthorhombic superstructure in very small areas of $Ca_{0.4}CoO_2$ as shown in fig. 2b, which should correspond with a local composition of around $Ca_{1/2}CoO_2$. Electron diffraction observations often demonstrate the presence of $3^{1/2}a \times 3^{1/2}a$ superstructure in $Ca_{0.3}CoO_2$ as shown in fig. 2c, which should correspond with a local composition of around $Ca_{1/3}CoO_2$. Actually, in the structural modulated point of view, the intercalated cation ordering in this layered $M_xCoO_2$ (M = Na, Sr, Ca; $0.33 \leq x \leq 0.5$) system can be described well by a structural modulation along the <110> direction with a modulation wave vector $q_1$ = <1/3, 1/3, 0> for x = 1/3 and $q_1$ = <1/4, 1/4, 0> for x = 1/2; this modulation shows up as multiple incommensurate states for x ranging from 0.3 to 0.5.

We now proceed to a detailed discussion of the structural evolution depending on the Ca content in $Ca_xCoO_2$ materials. Figure 3 shows a series of the [001] zone-axis diffraction patterns from several typical materials, illustrating remarkable structural changes as directly illustrated by the multiple arrangements of superstructure spots between the main diffraction spots of (000) and (110). In order to facilitate comparison, the diffraction patterns for the structural modulations are illustrated in the schematic representations as shown in the right column. In the present case, $q_1$ is alone the <110> direction of the hexagonal structure, and can be simplify written as $q_1 = q<110>$. It is demonstrated that in general the modulation vector decreases progressively with the increase of Ca content, which shows up a clear evolution from $3^{1/2}a \times 3^{1/2}a$ superstructure ($q = 1/3$) in $Ca_{0.3}CoO_2$, through numerous complex incommensurate phases to the well-defined orthorhombic superstructure ($q = 1/4$). Fig. 3(b) represents a [001] zone-axis diffraction pattern taken from $Ca_{0.3}CoO_2$, exhibiting the coexistence of two sets of weak reflection spots with $q = 0.33$ and $0.3$, respectively. Figure 3 (d) is a [001] zone-axis diffraction pattern taken from $Ca_{0.4}CoO_2$, showing coexistence of incommensurate structure $q = 0.27$ and the orthorhombic superstructure. It is also noted that the incommensurate wave vector might change slightly from one area to another even within one crystalline sample.

Another notable structural feature observed in our TEM investigations is the structural twinning arising from the intercalated cation ordering, which could give rise to interesting microstructure phenomenon as illustrated in fig. 4. There are three possible orientation states (crystallographically related variants) for cation/vacancy ordering within the hexagonal matrix. The coexistence of different variants leads to the formation of regular twining domains. In our TEM observations, electron diffraction directly demonstrated the coexistence of three variants within an individual crystal on the nanometer scale. Fig. 4 (a) and (b) show a [001] zone-axis experimental diffraction pattern and the schematic representation from a twinned crystal in the $Ca_{0.35}CoO_2$ materials. The superstructure reflections can be described by the coexistence of an incommensurate vector q = 0.28 and a superstructure q = 1/4. Fig. 4 (c) and (d) illustrate an experimental diffraction pattern and the schematic representation demonstrating the twining relationship with a relative 120º rotation. These three modulations yield three orthorhombic structures as marked by rectangles (in solid, dotted and doubled lines, respectively). High-resolution TEM observations indicated that the three twinned variants can coexist within one layer giving rise to twining domain structures and can also appear distinctively in different layers yielding 120º rotation in the cation ordered states.

A better and clear view of the atomic structure for the structural modulations and twinning has been obtained by the high-resolution TEM observations. Fig. 5(a) shows a [001] zone-axis HREM image for $Ca_{0.35}CoO_2$, showing the existence of twinned domains and clear twinning relationship. The well-defined 120° twinning is indicated by the broken lines. Superposition of three twinning variants can easily recognized in a number of areas, such as area "A", where the superstructure fringes show a hexagonal symmetry similar to the basic crystal structure. Figure 5(b) shows the result of a fast Fourier transform (FFT) of figure 5 (a); this pattern exhibits completely similar structural features as shown in the diffraction pattern of Figure 4 (d), demonstrating the presence of well-defined twining relationship for Ca-ordered states.

Fig. 5(c) and (d) show the [001] zone-axis HREM images for $Ca_{0.4}CoO_2$, illustrating the atomic structure correlates with the orientation anomaly and incommensurability. The orthorhombic supercell arising from the zigzag type of Ca ordering is clearly seen in both high-resolution TEM images and typically illustrated by a rectangle. Previously, incommensurate modulated structures with clear orientational and space anomalies have been analyzed based on the alternation of Na ordering in $Na_{0.5}CoO_2$ materials. Two structural models for explaining the observed incommensurability are shown in the insets of fig. 5 (c) and (d), respectively [13]. Careful analysis on the variation of atomic contrast in the TEM image suggests that incommensurability can be interpreted by an insertion of an extra vacancy plane as illustrated in the structural models. These models give the reasonable agreement with the observed HRTEM images.

As mentioned above, the structural properties in $Ca_xCoO_2$ materials become much complex for x<0.35. It is noted that a structural modulation with $q_2 = a^*/2$ as previously discussed in $Sr_{0.35}CoO_2$ [17] has been also observed in $Ca_xCoO_2$ ($0.15 \leq x \leq 0.35$) materials. Figure 6 (a) is the [010] zone-axis electron diffraction pattern taken from a crystal in the $Ca_{0.25}CoO_2$ sample, clearly showing the $q_2 = a^*/2$ reflection spots as indicated by arrows. Systematic structural analysis on the $Ca_xCoO_2$ ($0.15 \leq x \leq 0.35$) materials suggests that the $q_2$-modulation changes progressively with lowering Ca concentration. It appears as diffuse streaks along the c direction in $Ca_{0.35}CoO_2$, and becomes sharper with the decrease of x. i.e. the structural distortion firstly only appears in the c-axis direction and becomes visible in **a-b** plane with the further decrease of the cation content. Fig. 6 (b) displays an electron diffraction pattern taken from $Ca_{0.25}CoO_2$, demonstrating the coexistence of the cation ordering and structural modulations within the **a-b** plane.

In order to understand the microstructure properties of the $Na_{0.33}CoO_2 \cdot 1.3H_2O$ superconductor, we have also performed a series of TEM study focusing on the cation ordering and structural distortion. It is noted that the $Na_{0.33}CoO_2 \cdot yH_2O$ superconducting materials are easily damaged under electron beam due to the high mobility of Na atoms as well as the weak bonding of $H_2O$ molecules inside the crystal; We therefore performed TEM observations at temperature of around 100 K. It is found that radiation damage can be almost eliminated during our measurements. Fig. 6 (c) is a [001] zone-axis electron diffraction pattern of the superconducting hydrate, demonstrating the presence of $q_2$ = a*/2 within the a-b plane. These superstructure spots are sensitively dependent on the sample orientation, which could become weaker or stronger by slightly tilting the sample under observation. Hence, we propose that this modulation mainly arises from the structural distortion as we observed in $Sr_xCoO_2$ and $Ca_xCoO_2$ materials rather than Na ordering as reported in ref. 18. It is well known that the electric band structure can be greatly influence by the local structural distortion [19]. The superconducting behavior occurring in $Na_{0.33}CoO_2 \cdot yH_2O$ could be also influenced by this structural distortion. Further systematical theoretical work considering local structural distortion effect on electric structure in superconducting phase is still in progress.

## Conclusions

In summary, we have prepared a series of samples with nominal composition of $Ca_xCoO_2$ (0.15 ≤ x ≤ 0.40). TEM measurements revealed a rich variety of structural phenomena resulting from cation ordering, structural distortion and twinning domains. These notable structural features are found to change systematically along with the variation of intercalated Ca contents. Systematic analysis on the $Ca_xCoO_2$ ($0.15 \leq x \leq 0.40$) materials, in combination with previous structural results for $Na_xCoO_2$ ($0.33 \leq x \leq 0.80$) and $Sr_xCoO_2$ ($0.15 \leq x \leq 0.40$) systems, suggests that there are two well-defined cation ordered states corresponding respectively to the orthorhombic superstructure at around $x = 1/2$ and the $3^{1/2}a \times 3^{1/2}a$ superstructure at around $x = 1/3$. Complex ordered states, phase separation and structural twinning commonly appear in the materials with $0.33 < x < 0.5$. The TEM observations also reveal an additional periodic structural distortion with $\mathbf{q}_2 = a^* / 2$ in materials with a lower cation concentration ($x \leq 0.35$); this structural modulation also exists in the remarkable superconducting $Na_{0.33}CoO_2 \cdot 1.3H_2O$ material.

## Acknowledgments

We would like to thank Miss Y. Li for her assistance in preparing TEM samples and Dr. R.I. Walton for fruitful discussions and help during manuscript preparation. The work reported here is supported by 'National Natural Foundation' and 'Outstanding Youth Fund' of China.

## Reference and Notes

1. K. Mizushima, P.C. Jones, P.J. Wiseman, J.B. Goodenough, *Mater. Res. Bull.* **15**, 783 (1980)

2. W. Koshibae, K. Tsutsui, S. Maekawa, *Phys. Rev. B*, **62**, 6869 (2000).

3. K. Takada, H. Sakurai, E. Takayama-Muromachi, F. Izumi, R.A. Dilanian, T. Sasaki, *Nature*, **422**, 53 (2003).

4. J.N. Reimers, J. R. Dahn, *J. Electrochem. Soc.* **139**, 2091 (1992).

5. M.L. Foo, Y.Y. Wang, S. Watauchi, H.W. Zandbergen, T. He, R.J. Cava, N.P. Ong, *Phys. Rev. Lett*. **92,** 247001 (2004).

6. P.H. Zhang, R.B. Capaz, M.L. Cohen, S.G. Louie, *Phys. Rev. B*, **71**, 153102 (2005).

7. A. Van der Ven, M.K. Aydinol, G. Ceder, *Phys. Rev. B*, **58**, 2975 (1998).

8. Y. Shao-Horn, S. Levasseur, F. Weill, C. Delmas, *J. Electrochem. Soc.* **150**, A366 (2003).

9. J. P. Pérès, F. Weill, C. Delmas, *Solid State Ionics,* **116,** 19 (1999).

10. H.X. Yang, Y.G. Shi, C.J. Nie, D. Wu, L.X. Yang, *Materials Chemistry and Physics* (in press)

11. Q. Huang, M.L. Foo, J.W. Lynn, H.W. Zandbergen, G. Lawes, *J. Phys.: Condens. Matter*, **16**, 5803 (2004).

12. H.W. Zandbergen, M. Foo, Q. Xu, V. Kumar, R.J. Cava, *Phys. Rev. B*, **70,** 024101 (2004).

13. H.X. Yang, C.J. Nie, Y.G. Shi, H.C. Yu, S. Ding, Y.L. Liu, D. Wu, N.L. Wang, J.Q. Li, *Solid state Communications*, **134**, 403 (2005).

14. Y.G. Shi, J.Q. Li, H.C. Yu, Y.Q. Zhou, H.R. Zhang, C. Dong, *Supercon. Sci. Tech*. **17**, 42 (2004).

15. B.L. Cushing, J.B. Wiley, *J. Solid. State. Chem.* **141**, 385 (1998).

16. R. Ishikawa, Y. Ono, Y. Miyazaki, T. Kajitani, *Japn. J. Appl. Phys.* **41**, L337 (2002).

17. H.X. Yang, Y.G. Shi, Y.Q. Guo, X. Liu, R.J. Xiao, J.L. Luo, J.Q. Li, *cond-mat/0503136.*

18. D.N. Argyrious, P.G. Radaelli, C.J. Milnet; N. Aliouane, L.C. Chapon, A. Chemseddinet, J. Veira, S. Cox, N.D. Mathur, P.A. Midgley, *http://www.hmi.de/people/argyriou/recent_publications.html.*

19. J.Q. Li, *J. App. Phys.* **90**, 637 (2001).

**Table 1** Lattice parameters of $M_xCoO_2$ (M = Na, Ca). The structural Rietveld refinements have carried out by using the conventional $P6_3/mmc$ symmetry.

| samples | a(Å) | c(Å) |
|---|---|---|
| $Ca_{0.15}CoO_2$ | 2.827 | 10.893 |
| $Ca_{0.25}CoO_2$ | 2.811 | 10.887 |
| $Ca_{0.35}CoO_2$ | 2.815 | 10.886 |
| $Ca_{0.4}CoO_2$ | 2.819 | 10.881 |
| $Na_{0.5}CoO_2$ | 2.824 | 11.143 |
| $Na_{0.70}CoO_2$ | 2.843 | 10.804 |

**Figure captions**

Fig. 1 XRD patterns of $Ca_{0.25}CoO_2$ and $Na_{0.5}CoO_2$, indicating that the orthorhombic structural distortion in $Na_{0.5}CoO_2$ which arises from the Na zigzag ordering; $Ca_{0.25}CoO_2$ has a hexagonal basic structure.

Fig. 2 (a) A brief phase diagram illustrating the presence of ordered structures depending on cation content. Electron diffraction patterns appear in some area of (b) $Ca_{0.4}CoO_2$ and (c) $Ca_{0.3}CoO_2$ along the [001]-zone axis directions, representing two well-defined ordered structures caused by cation and vacancy ordering in the $M_xCoO_2$ compounds. (d) Dark-field TEM image from a satellite spot showing the phase separation.

Fig. 3 (a) [001] zone-axis experimental diffraction pattern and (b) the schematic representations for samples with different Ca contents. The small black dots and circles represent, respectively, the commensurate superstructure reflections and incommensurate superstructure reflections.

Fig. 4 (a) [001] zone-axis diffraction and (b) the schematic illustration taken from a crystal in the $Ca_{0.35}CoO_2$ material. The superstructure spots (circles) and incommensurate satellites (*) are indicated by two sets of arrows.

(c) [001] zone-axis diffraction pattern taken from a twinned crystal in $Ca_{0.35}CoO_2$ and (d) the schematic illustration clearly exhibits the twinning relationship. The three resultant orthorhombic cells are marked by rectangles (in solid, dotted and doubled lines).

Fig. 5 (a) [001] zone-axis HREM image for $Ca_{0.35}CoO_2$, showing the presence of complex superstructures and twining domains. (b) Fast Fourier transform pattern for the HREM image of fig. 5(a). HREM images for the incommensurate modulations with (c) spatial anomaly and (d) orientation anomaly, structural models are shown in the insets, Co atoms are shown as gray dots and Ca atoms are shown as dark dots.

Fig. 6 Electron diffraction pattern showing the presence of the superstructure in $Ca_{0.25}CoO_2$ (a) taken along [010] zone axis direction and (b) taken along [001] zone axis direction. (c) Electron diffraction patterns of $Na_{0.33}CoO_2 \cdot 1.3H_2O$ superconductor, showing the presence $\mathbf{q_2} = a^*/2$.

**Figure 1**

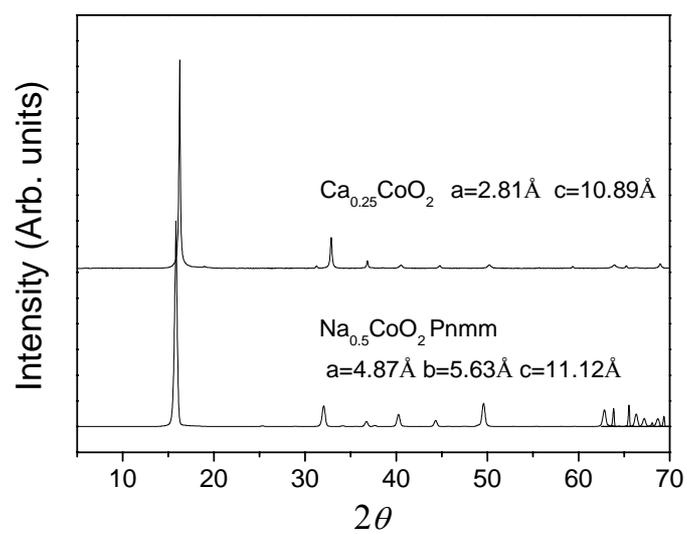

**Figure 2**

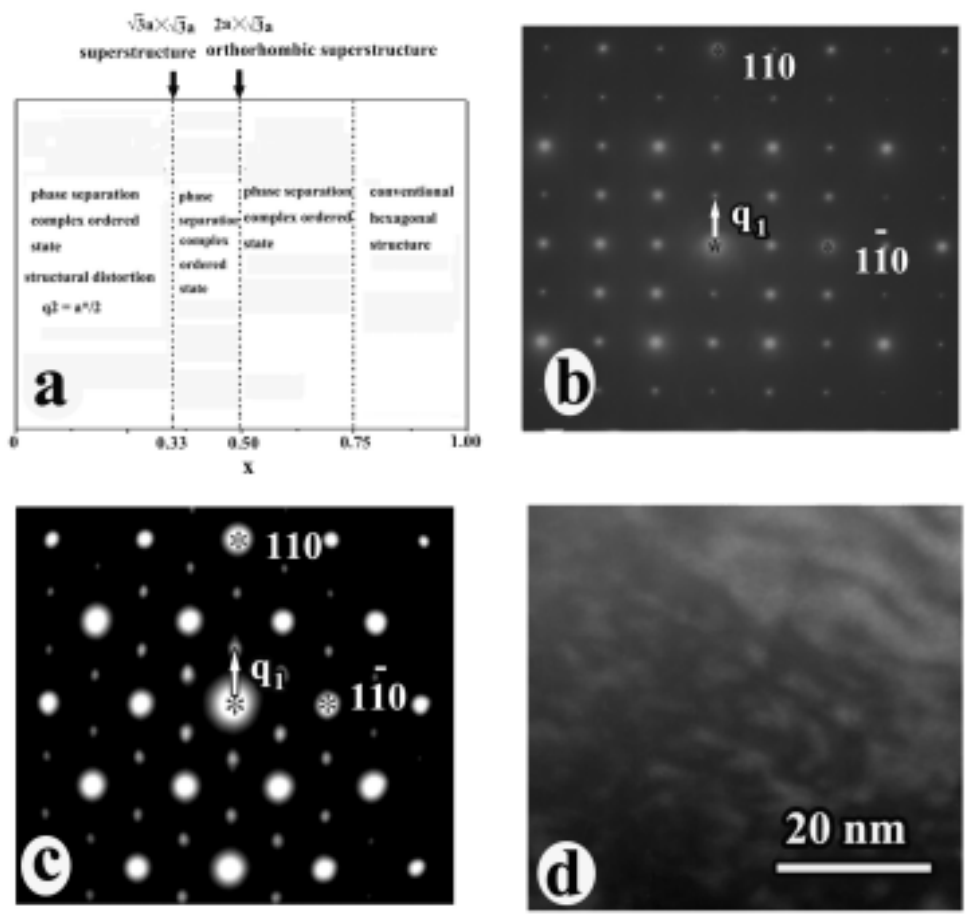

**Figure 3**

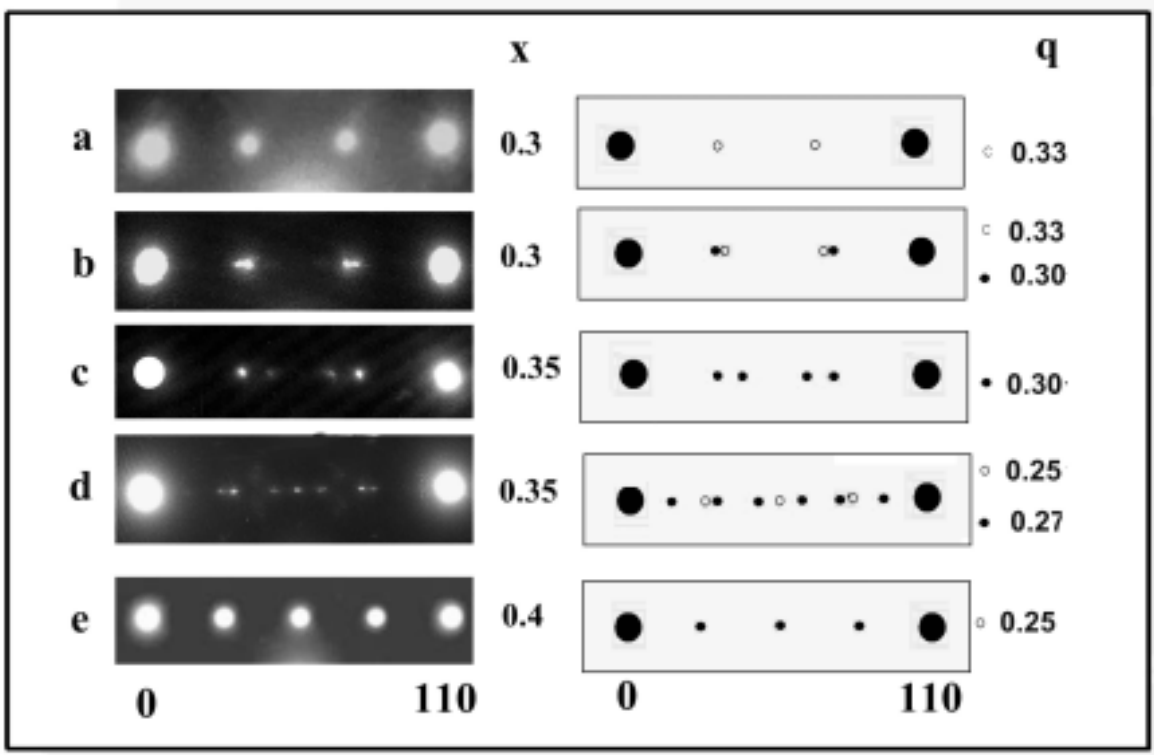

**Figure 4**

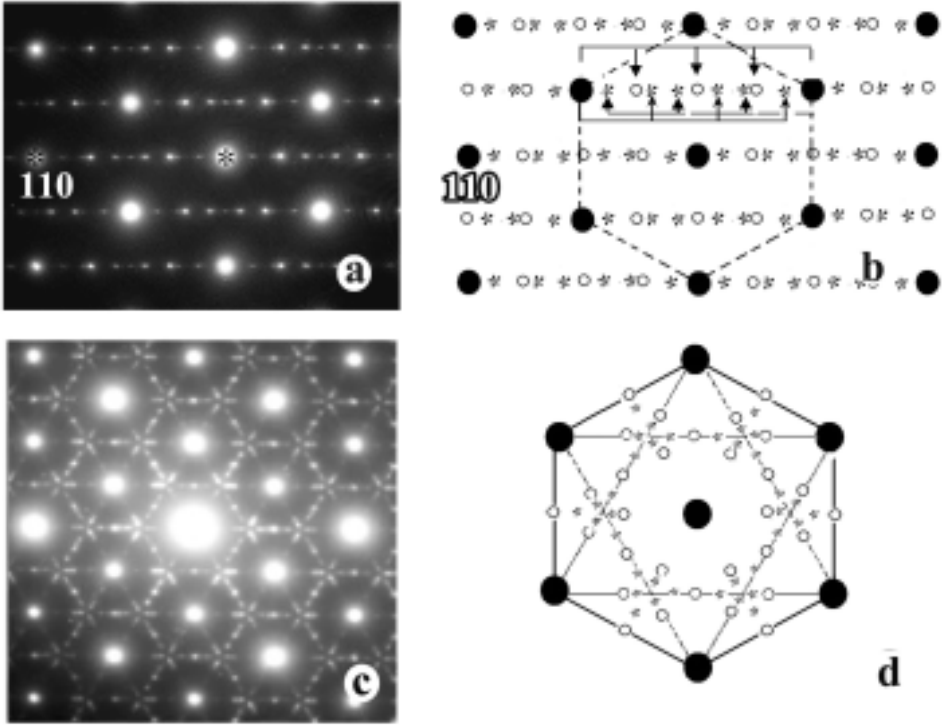

**Figure 5**

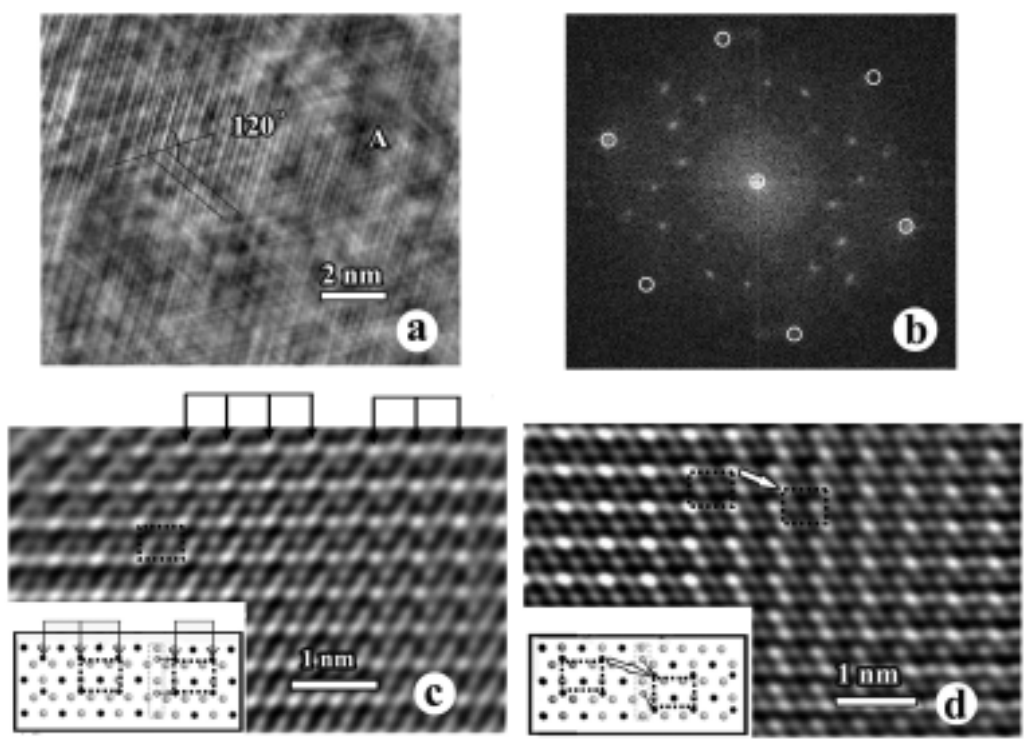

**Figure 6**

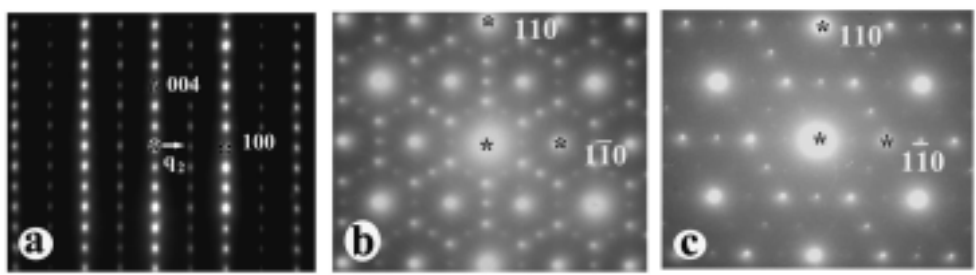